\documentclass[fleqn,usenatbib,letters]{mnras}

\usepackage{newtxtext,newtxmath}

\usepackage[T1]{fontenc}
\usepackage{ae,aecompl}

\usepackage{graphicx}	
\usepackage[export]{adjustbox}
\usepackage{amsmath}	

\usepackage{color}

\newcommand{\ec}[1]{{\color{red} #1}}

\defcitealias{2021MNRAS.507..971C}{Paper~I}

\title[LOFAR detection of G116.6-26.1]{LOFAR detection of faint radio emission from the supernova remnant SRGeJ0023+3625=G116.6-26.1: probing the Milky Way synchrotron halo.}

\author[Churazov et al.]{E.M.~Churazov,$^{1,2}$ I.I.~Khabibullin,$^{3,2,1}$ A.M.~Bykov,$^4$ N.N.~Chugai,$^5$ R.A.~Sunyaev,$^{1,2}$ and I.I.~Zinchenko$^{6}$
\\
\\
$^1$~Space Research Institute (IKI), Profsoyuznaya 84/32, Moscow 117997, Russia \\
$^2$~Max Planck Institute for Astrophysics, Karl-Schwarzschild-Str. 1, D-85741 Garching, Germany  \\
$^3$~Universitäts-Sternwarte, Fakultät für Physik, Ludwig-Maximilians-Universität München, Scheinerstr.1, 81679 München, Germany \\
$^4$~Ioffe Institute, 26 Politekhnicheskaya str., St. Petersburg 194021, Russia \\
$^5$~Institute of Astronomy, Russian Academy of Sciences, 48 Pyatnitskaya str., Moscow 119017, Russia  \\
$^6$~Institute of Applied Physics of the Russian Academy of Sciences, 46 Ul'yanov~str., Nizhny Novgorod 603950, Russia. \\
}

\begin{document}
\label{firstpage}
\pagerange{\pageref{firstpage}--\pageref{lastpage}}
\maketitle

\begin{abstract}
A supernova remnant (SNR) candidate SRGe~J0023+3625 = G116.6-26.1 was recently discovered in the  \textit{SRG}/eROSITA all-sky X-ray survey. This large ($\sim 4$ deg in diameter) SNR candidate lacks prominent counterparts in other bands. Here we report detection of radio emission from G116.6-26.1 in the LOFAR Two-metre Sky Survey (LoTTS-DR2). Radio images show a shell-like structure coincident with the X-ray boundary of the SNR. The measured surface brightness of radio emission from this SNR is very low. Extrapolation of the observed surface brightness to 1~GHz   places G116.6-26.1 well below other objects in the $\Sigma-D$ diagram.  We argue that the detected radio flux might be consistent with the minimal level expected in the
van der Laan adiabatic compression model, provided that the volume emissivity of the halo gas in the LOFAR band is $\sim 10^{-42}\,{\rm Wm^{-3}Hz^{-1} sr^{-1}}$. If true, this SNR can be considered as a prototypical example of an evolved SNR in the Milky Way halo. In the X-ray and radio bands, such SNRs can be used as  probes of thermal and non-thermal components constituting the Milky Way halo.   
\end{abstract}





\section{Introduction}
\begin{figure*}
\centering
\includegraphics[angle=0,trim=0cm 0cm 0cm 0cm,width=1.95\columnwidth]{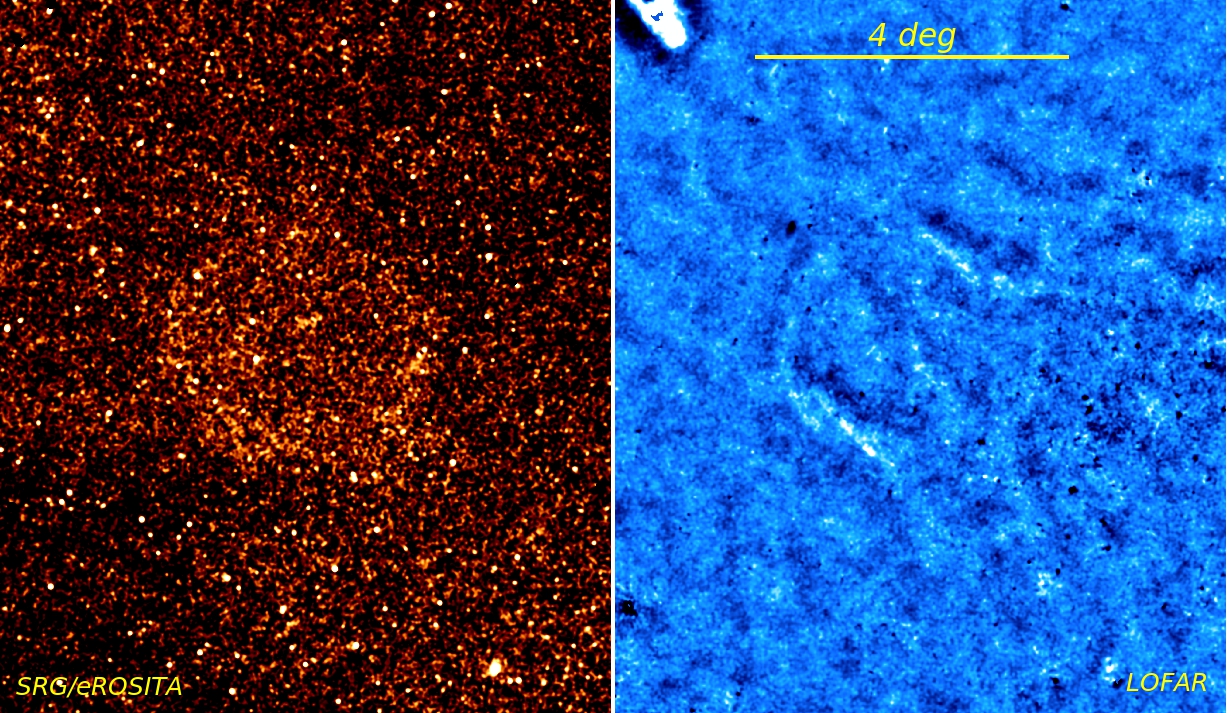}
\caption{Side by side comparison of the X-ray (0.5-0.7~keV; eROSITA) and radio (144~MHz; LOFAR) images of the SRGe~J0023+3625 field (equatorial orientation). Bright compact sources in the LOFAR image have been masked; the image was convolved with $\sigma=30"$ Gaussian. The bright object in the upper left corner is M31 (Andromeda galaxy).
}
\label{fig:image}
\end{figure*}

A supernova remnant (SNR) candidate SRGe~J0023+3625 = G116.6-26.1 was found \citep[][hereafter, Paper I]{2021MNRAS.507..971C} in the  \textit{SRG}/eROSITA all-sky survey. 
The \textit{SRG} X-ray observatory \citep{2021A&A...656A.132S} with the two wide-angle grazing-incidence X-ray telescopes, eROSITA \citep[0.3-10 keV,][]{2021A&A...647A...1P} and Mikhail Pavlinsky ART-XC telescope \citep[4-30 keV,][]{2021A&A...650A..42P} started its all-sky X-ray survey on December 13, 2019. G116.6-26.1 features a large angular extent ($\sim 4^\circ$ in diameter, see the left panel of Fig.~\ref{fig:image}), nearly circular shape, and X-ray spectrum dominated by emission lines of helium- and hydrogen-like oxygen. Given the relatively high Galactic latitude of the source, $b\approx-26^\circ$, this SNR is plausibly associated with a Type Ia supernova that exploded some 40 000 yr ago in the low density gas of the Milky Way halo. G116.6-26.1 lacks bright counterparts of similar extent at other wavelengths, in particular, we didn't find clear evidence of radio features in the 408~MHz survey \citep{Haslam82}. The hope was that higher sensitivity of the LOFAR survey at lower frequencies might be able to find the radio counterpart and confirm the SNR nature of this object.

\begin{figure}
\centering
\includegraphics[width=\columnwidth,bb=60 200 560 680 ]{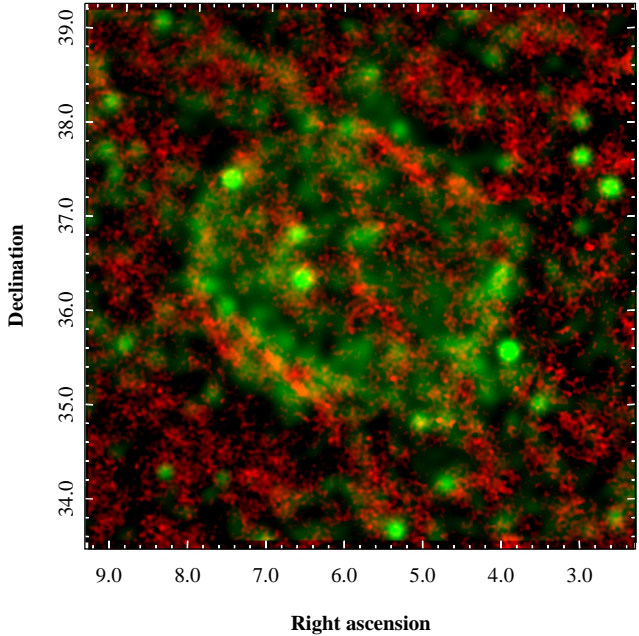}
\caption{A composite image showing surface brightness of
the radio emission at 144~MHz after masking of the point sources (in red) and diffuse X-ray emission in the 0.5-0.7 keV band smoothed with a $\sigma=$4' Gaussian filter (in green).
}
\label{fig:image_rg}
\end{figure}

\section{Comparison of X-ray and radio data on G116.6-26.1}
The recently released LoTSS-DR2 data \citep{2022arXiv220211733S} in the 120-168 MHz cover 27\% of the northern sky. In particular, the data cover the G116.6-26.1 field. We used the $10\times 10$ degrees field, centered at the candidate object. Multiple bright radio sources dominating the LOFAR image have been masked and the resulting image smoothed with the $\sigma=30"$ Gaussian filter (see the right panel of Fig.~\ref{fig:image}). The comparison of the X-ray and radio images immediately shows the edge-brightened structures in the LOFAR image coincident with the edges of the X-ray image (see also Fig.~\ref{fig:image_rg}). Clearly, the amplitude of the structures in the LOFAR image is comparable with the fluctuations seen everywhere in the image at comparable spatial scales. However, in combination with the X-ray data, it is clear that the "edges" of G116.6-26.1 are indeed present in the radio data. Noteworthy, the azimuthal distribution of the radio brightness gives a hint of a bi-lateral morphology. If confirmed by future high sensitivity data, this would indicate a
coherence length for the halo magnetic field of at least $\sim200$\,pc (see Discussion).

\begin{figure}
\centering
\includegraphics[angle=0,trim=1cm 5cm 0cm 4cm,width=0.95\columnwidth]{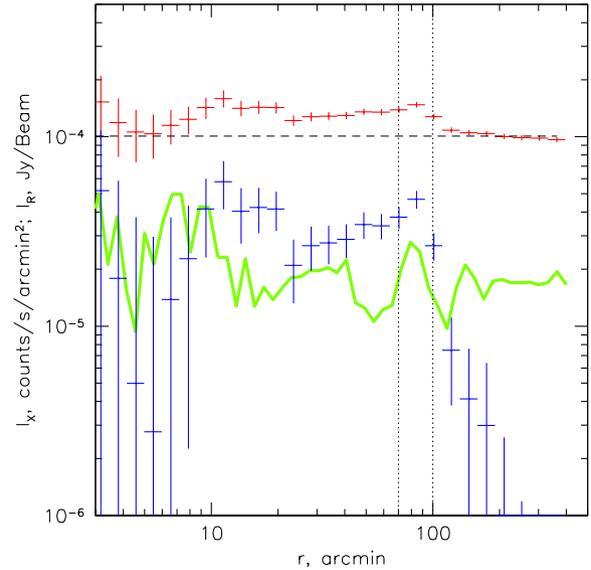}
\caption{Comparison of radial profiles of the SNR candidate SRGe~J0023+3625 in the X-ray (eROSITA; red, blue) and radio (LOFAR; green) bands. The profile in the 0.5-0.7 keV band \citepalias[same as in][see their fig.2]{2021MNRAS.507..971C} is in units of ${\rm counts\,s^{-1}\,arcmin^2}$ per one eROSITA telescope. The red points show the total sky surface brightness, while the blue points show the same data once the estimated sky background level is subtracted. For comparison, the green curve shows the radial profile extracted from the LOFAR image (after masking bright point sources). Due to the limited $uv$ coverage of the LOFAR DR2 data, much of the extended emission (if any) might be filtered out. However, sharp features should be preserved in the data. In particular, the peak at $100'$ is clearly present in both X-ray and radio data. 
}
\label{fig:radial}
\end{figure}

Fig.~\ref{fig:radial} shows radial profiles of G116.6-26.1 in the X-ray (red and blue) and radio (green) bands. Apart from the complicated structure in the central $\sim 10'$, there is a prominent peak at $\sim 100'$, which coincides with the peak of the X-ray surface brightness. The excess surface brightness in the radio band 
is $\sim 10^{-5}\,{\rm Jy/beam}$. 


According to the description of the LoTSS-DR2 data, the baselines shorter than 0.1~km are not used when making images \citep{2022arXiv220211733S} and, therefore, large-scale diffuse structures can be significantly suppressed and the observed surface might be much lower than the true value. We estimate the characteristic size of the strongly suppressed scales 
as $S_{max}\approx 0.6 c/(\nu_{max}D)\approx 36'$ \citep[see eq.3.28 in][]{almahandbook}, using $\nu_{max}=168\,{\rm MHz}$ and $D=0.1\,{\rm km}$. Assuming that $S_{max}$ can be treated as the FHWM of the Gaussian kernel, we estimated the drop of the surface brightness associated with high-pass filtering for two very simple models. For a flat disk with the radius of $r=100'$, the filtered image looks like a ring, with the maximal surface brightness of $\sim0.5$ of the initial value at the outer radius. Inside the disk, the surface brightness decreases towards the center: it drops by a factor of $~2$ at  $r\sim 90'$ and then decreases very rapidly at smaller radii. For a $10'$-wide ring ($r=[90'-100']$), the surface brightness drops to $\sim 70-75$\% of the initial value. Therefore, it is not easy to distinguish the disk with a sharp edge and the ring models based on the LoTSS-DR2 images. The peak surface brightness is suppressed in the flat disk and ring models by a factor $\eta\sim 0.5-0.7$. However, for the mean surface brightness of the $r=100'$ disk, the suppression factor is larger $\eta\sim 0.1$. 


\section{Discussion}
The detection of radio signal co-spatial with SRGe~J0023+3625 confirms the identification of this object as SNR. The LOFAR detection is robust (see Fig.~\ref{fig:image}), but the surface brightness is very low (see Fig.~\ref{fig:radial}). It is interesting to compare the latter property with other SNRs. We do this by estimating the position of G116.6--26.1 in the so-called $\Sigma-D$ diagram \citep[e.g.][]{1976MNRAS.174..267C,Green84,2005A&A...435..437U,2019SerAJ.199...23S}, where $\Sigma$ is the surface brightness at $1\,{\rm GHz}$ and $D$ is the diameter of the SNR, respectively. 

\begin{figure}
\centering
\includegraphics[bb= 50 170 560 680,width=0.95\columnwidth]{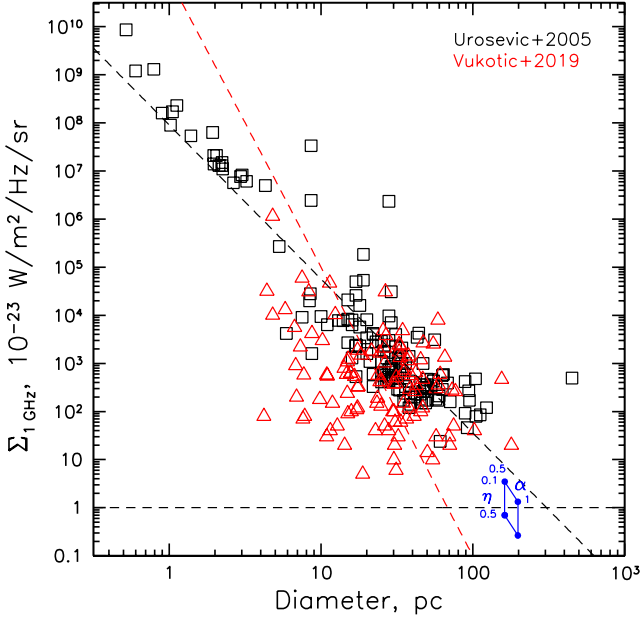}
\caption{Position of G116.6--26.1 (blue points) on the $\Sigma-D$ diagram for extragalactic  \citep[][]{2005A&A...435..437U} and Galactic \citep[][]{2019SerAJ.199...23S} supernova remnants, depending on the assumed spectral slope $\alpha$, filtering suppression factor $\eta$ and the line-of-sight distance. The dashed lines show corresponding power-law scaling relations derived for these samples. }
\label{fig:sigmad}
\end{figure}

Since the large scale diffuse emission is filtered out in the LOFAR images, we use the peak surface brightness instead of the mean and extrapolate from $\nu=144~{\rm MHz}$ to $1\,{\rm GHz}$ assuming a power law spectrum with spectral index $\alpha=0.5$
\begin{eqnarray}
\Sigma_{1\,{\rm GHz}}\approx I_{peak}\times \frac{1}{\eta} \left (\frac{1\,{\rm GHz}}{\nu} \right )^{-\alpha}\approx \nonumber \\
7\,10^{-24} \left( \frac{I_{peak}}{10\,{\rm \mu Jy/beam}} \right ) \left( \frac{\eta}{0.5} \right )^{-1} \, {\rm W \, m^{-2}Hz^{-1} sr^{-1}},
\label{eq:sigma}
\end{eqnarray}
where the beam size of $20''$ was used for the low resolution version of the LoTSS-DR2 images. The corresponding values of $\Sigma_{1\,{\rm GHz}}$ are shown in Fig.~\ref{fig:sigmad} (blue points) together with the data from two SNR samples. For  G116.6--26.1, the uncertainties (apart from the noisy data) come from the uncertainty in the adopted spectral index $\alpha$, the distance to the SNR and filtering attenuation factor $\eta$. But even allowing for uncertainties, it is clear that G116.6--26.1 is located at the very bottom of the $\Sigma-D$ diagram that is poorly covered by the existing surveys.

The standard assumption is that the radio emission from young SNRs is powered by the DSA mechanism \citep[see e.g.][]{Reynolds81,Helder12}. During the late stages of an SNR evolution, the surface brightness gradually goes down as the size of the SNR increases and the expansion velocity decreases \citep[e.g.][]{BV04,2006A&A...459..519A}. This qualitatively explains the $\Sigma-D$ diagram, albeit with large uncertainties. Broadly, the parameters of G116.6--26.1 are consistent with the extension of this diagram. However, given that the surface brightness of G116.6--26.1 in the LOFAR band is very low, it is interesting to consider yet another possibility, namely adiabatic compression of the existing population of relativistic electrons in the Milky Way halo. For instance, \cite{1975MNRAS.171..243W,2013ApJ...776...42S} postulate the existence of a very extended (10-15~kpc) halo around the Milky Way, which can explain some of the sky brightness in the radio band \citep[see, however][for counter arguments]{2010MNRAS.409.1172S}.

\begin{figure}
\centering
\includegraphics[angle=0,trim=1cm 5cm 0cm 4cm,width=0.95\columnwidth]{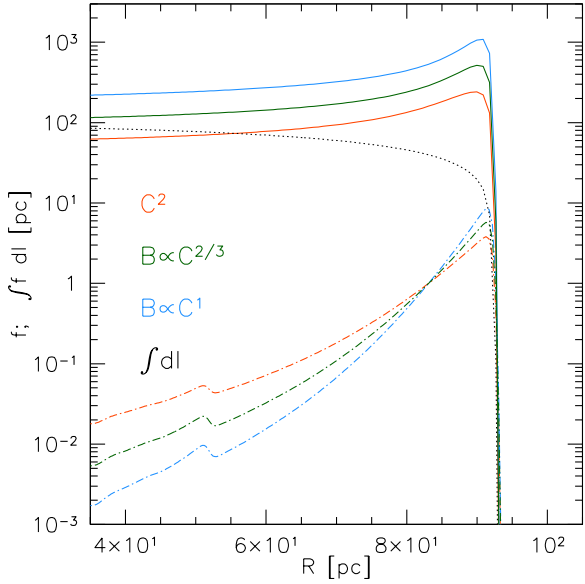}
\caption{"Boost factor" $f$ of the volume emissivity (dashed-dotted lines) and the surface brightness profiles $\int f dl$ (solid lines), associated with the adiabatic gas compression in the van der Laan model. The density ratio is taken from the illustrative simulation of the Type Ia explosion in the low-density and hot gas of the Milky Way halo. The red curve is for the emission measure (square of the density). The blue and green curves assume different magnetic field amplification due to compression.  For comparison, the black dotted line shows the effective length of the l.o.s. inside the volume encompassed by the shock, i.e. $\int dl$.
}
\label{fig:proj}
\end{figure}

 Since we are interested only in the adiabatic compression (other gains or losses are neglected), the problem is fully specified by the volume emissivity of the gas around the SNR in the radio band $\varepsilon_{r,0}$, the compression/rarefaction of the gas, and magnetic field in the shock downstream. The model was developed by \citet{1962MNRAS.124..125V} for the extended evolved SNRs like the Cygnus Loop and IC 443. The origin of the non-thermal emission from evolved SNRs with relatively weak shocks may differ from young objects like e.g. SN 1006.  In strong forward shocks in young SNRs, the Diffusive Shock Acceleration (DSA) mechanism is considered to be efficient \ec{enough} to accelerate electrons to TeV regime energies and provide super-adiabatic amplification of turbulent magnetic fields by cosmic ray driven instabilities   \citep[see e.g.][]{Helder12,Schure_Bell12}. In general, in the middle-aged and old SNRs, both mechanisms (DSA and adiabatic compression) can contribute to the observed radio emission depending on the shock characteristics and the ambient matter conditions. We consider here a 'minimal scenario' with the pure adiabatic boost of the synchrotron emissivity by van der Laan's mechanism.  For a power law distribution of relativistic particles with slope $p$, the boost of the volume emissivity in the radio band due to adiabatic compression with the density ratio $C=\rho_d/\rho_u$ (where $\rho_d$ and $\rho_u$ are the downstream and upstream densities, respectively) can be written as
\begin{eqnarray}
f_{r}=C_b^\frac{p+1}{2} C^\frac{p-1}{3} C,
\label{eq:fboost}
\end{eqnarray}
where the $C_b$ reflects the change of magnetic field due to compression, the second term accounts for the change of the particles' Lorentz factor $\gamma/\gamma_0=C^{1/3}$, and the last term stands for the increase of particles density. We used $p=2\alpha+1=2.4$, motivated by the spectral index $\alpha=0.7$ found by MeerKat \citep{2022MNRAS.509.4923I}. For $C_b$ we consider two cases: $C_b=C$ and $C_b=C^{2/3}$. The former corresponds to the case when the magnetic field is along the shock front (can be relevant for selected patches of the shock surface), while the latter assumes that the field is random and behaves like the gas with adiabatic index $4/3$. In all cases, an isotropic distribution function of relativistic electrons is assumed. Fig.~\ref{fig:proj} shows the radial profiles of the volume emissivity boost factors for these two cases (dashed lines). Here we use $C$ from the simple hydrodynamic simulations of G116.6--26.1 \citepalias{2021MNRAS.507..971C}.  For comparison, the boost in the X-ray volume emissivity $f_x=C^2$ is also shown\footnote{Note that this factor accounts only for the density change. Another important factor for this SNR is the emissivity boost due to non-equilibrium ionization \citepalias[][]{2021MNRAS.507..971C}, which we discuss bellow}. From the observational point of view, a more important is the boost factor $f_{r}$ integrated along the line of sight $I(R)=\int f_{r} dl$ (within the spherical volume encompassed by the shock). This quantity, which characterizes the surface brightness increase, is also shown in Fig.~\ref{fig:proj} with colored solid lines. For comparison, the black solid line shows the same quantity for $f_{r}=1$. In this case, the surface brightness profile $I_1(R)$ simply reflects the length of the l.o.s. inside the shock volume. With these definitions the change in the surface brightness associated with the presence of the shock is 
\begin{eqnarray}
\delta I(R)=\varepsilon_{r,0}\times \left [ I(R)-I_1(R) \right ],
\label{eq:dir}
\end{eqnarray}
where $\varepsilon_{r,0}$ is the volume emissivity of unperturbed gas.
Combining Eq.~\ref{eq:sigma} (for 144~GHz) and the peak values for two curves (for $C_B=C$ and $C_b=C^{2/3}$) from Fig.~\ref{fig:proj}, we estimate the required emissivity
\begin{eqnarray}
\varepsilon_{r,0}\sim 0.6-1.2\, 10^{-42} \left( \frac{\eta}{0.5} \right )^{-1} \,{\rm Wm^{-3}Hz^{-1} sr^{-1}}
\end{eqnarray}
Interestingly, $\varepsilon_{r,0}$ turns out to be close to values suggested in  \cite{1975MNRAS.171..243W,2013ApJ...776...42S} in the frame of the extended halo model. While this exercise does not prove that the halo model is correct, it illustrates that using SNRs in the halo, one can get interesting constraints on the emissivity of the medium.  

Therefore, even if DSA is not operating at the later stage of SNR evolution, the adiabatic compression is fully sufficient. This argument can of course be reversed and the observed surface brightness can be used as an upper limit on the DSA efficiency or  $\varepsilon_{r,0}$. 
Note here that if SNR is expanding in a locally homogeneous regular magnetic field then the adiabatic compression mechanism discussed above would predict an anisotropic image of the radio emission. Indeed from the observations of the galactic foreground rotation-measure distribution \citet{Beck16} derived a correlation length of the magnetic field to be about 220 pc (with 50 pc 5$\sigma$ lower limit). Therefore one can expect a relatively homogeneous ambient magnetic field at the scale size of the SNR. The boosting factor of the synchrotron emissivity in the \cite{1962MNRAS.124..125V} adiabatic compression mechanism  is maximal for the parts of the shock expanding transverse to the local field direction. On the other hand, the diffusive acceleration mechanism could also produce  anisotropy.
The model has a minimal number of parameters. Since the ambient synchrotron emissivity of the halo is likely rather homogeneous at a kpc scale height therefore SNR G116.6-26.1 can be considered as a radio test bed  for a typical SNR in the Milky Way halo. Recently, \citet{Raymond20} discovered in the optical band an evolved Type Ia SNR G70.0–21.5 of the estimated age 90,000 yrs located in the halo at a distance of about 1 kpc. Radio and X-ray studies of the SNR may provide another test of the simple model. The sizes, properties, and origins of radio halos in spiral galaxies is a subject of active studies \citep[see e.g.][]{beck15,krause19} and new very sensitive instruments like MeerKat and  SKA are really needed to resolve the issue. On the other hand, an indirect approach based on a model interpretation of the galactic cosmic ray Be/B ratio measured by AMS-02 was used to constrain the cosmic ray halo size as $5^{+3}_{-2}$ kpc at 68\% confidence level \citep{AMS02_halo20}. The size of the galactic cosmic ray halo is connected to the magnetic field scale height.     

\begin{figure}
\centering
\includegraphics[angle=0,trim=1cm 5cm 0cm 4cm,width=0.95\columnwidth]{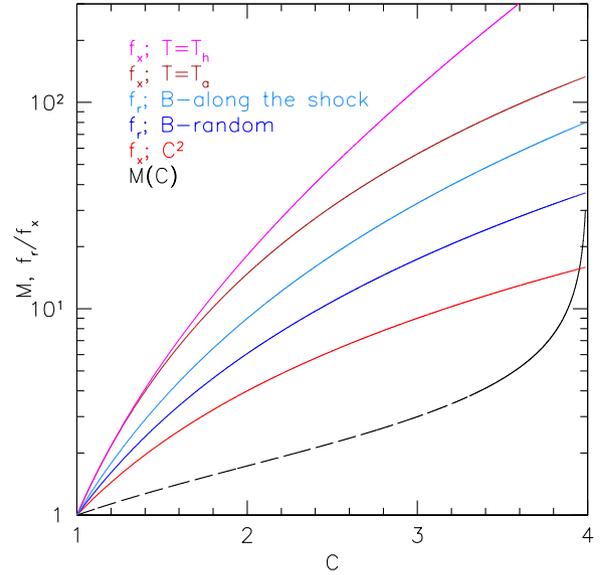}
\caption{"Boost factors" in the volume emissivity as a function of the compression factor $C$. The black dashed in the Mach $M(C)$ number for the gas with adiabatic index $5/3$. For the X-ray volume emissivity, the boost factor $f_x$ is driven by the square of the density ($C^2$; the red curve) and, possibly, by the non-equilibrium ionization (NEI) effects (magenta and brown curves). It is likely that the effect of the NEI is overestimated and the correct value is somewhere between the red and magenta curves. For the volume emissivity in the radio band, the value $f_r$ in the van der Laan model depends on the orientation of the magnetic field with respect to the shock front (two bluish curves). The main conclusion one can draw from this plot is that for $M\sim3$, both X-ray and radio volume emissivities are boosted by a factor of order 10. 
}
\label{fig:fxfr}
\end{figure}

It is interesting that in the scenario when the adiabatic compression powers the synchrotron emission and the radiative losses of the shock heated gas can be neglected (the case of G116.6--26.1), the X-ray and radio bands are tightly related, Indeed, both the radio and X-ray volume emissivities can be expressed via upstream emissivities $\varepsilon_{0,r}$ and $\varepsilon_{0,X}$, respectively and the compression ratio $C$. For the X-ray band, the boost factor is 
\begin{eqnarray}
f_x=C^{2} N(C),
\label{eq:fx}
\end{eqnarray}
where $N(C)$ describes the impact of non-equilibrium ionization (NEI) on the X-ray emissivity of the gas.  To the first approximation, the NEI factor can be written as 
\begin{eqnarray}
N(C)\sim e^{-E/kT+E/kT_0},
\end{eqnarray}
where $E$ is the observed energy range, $T_0$ and $T$ are electron temperatures upstream and downstream, respectively. This expression assumes that the ionization state does not change much after the passage of the shock (due to an extremely low density of the gas), but electrons are now hotter and excite relevant transitions more efficiently. This is, of course, the maximal possible boost factor. In real systems, it is expected to be smaller. We adopt
$T_0\sim 2 10^6\,{\rm K}$ as the characteristic temperature of the hot gas in the Milky Way halo, while $T$ is the function of the compression factor $C$. We consider two limiting scenarios for $T$: pure adiabatic heating of electrons, i.e. $T=T_a=T_0 C^{2/3}$ and higher value $T=T_h$ that follows from the Rankine-Hugoniot condition for the gas with the adiabatic index $5/3$. The true electron temperature is somewhere in between $T_a$ and $T_h$. The boost factors $f_r$ and $f_x$ are shown in Fig.~\ref{fig:fxfr}. While the scatter is large, it is plausible that both X-ray and radio volume emissivities are enhanced by a comparable factor $\sim 10$.

The number of SNRs in the Milky Way halo (presumably remnants of type Ia SN) is small. They are embedded into the hot low-density gas of the halo, and, therefore, are expected to be X-ray faint. Departures from ionization equilibrium boost their X-ray fluxes and improve the chances of being detected in the sensitive all-sky surveys, such as the \textit{SRG}/eROSITA survey. Similarly, the presence of extended synchrotron halo implies that even in the absence of DSA mechanism, the adiabatic compression of the gas boosts the emissivity in the low-frequency band by the factor of $\sim 10$ and sets the minimum level of the radio surface brightness, within reach of the LOFAR or MeerKat surveys. If the DSA mechanism is efficient, the surface brightness might be be even higher. G116.6--26.1, which is now detected in both the X-ray and radio bands, might be an example of such an object. The measured fluxes provide a new probe on the properties of the thermal gas in the halo (at the location of the SNR), as well as the non-thermal component. In principle, radio polarization observations may help to distinguish between the adiabatic compression and DSA mechanisms.  

Another extended (about 4.4 degrees in diameter) high latitude galactic SNR  G249.5+24.5 (Hoinga) was recently discovered by \textit{SRG}/eROSITA observatory \citet{Becker21}. The authors constrained the distance D to be within the range of 0.45 < D < 1.2 kpc. The estimated local number density 0.42–0.16 cm$^{-3}$ and the radio flux obtained below 2 GHz are higher than that in the case of G116.6-26.1 discussed above. The Hoinga likely has the lower height than G116.6-26.1 in the disk-halo transition layer. The radio spectral index obtained for the Hoinga was 0.69$\pm 0.08$ and linear polarization is detected from the brightest parts of the shell with high compression. Both these characteristics are consistent with the adiabatic compression scenario discussed above while the possible role of diffusive shock acceleration mechanism requires more studies. In this respect, it would be especially interesting to determine the radio emission indexes and polarization of yet another high-latitude SNR G70.0–21.5 \citep{Raymond20}. 



\section{Conclusions}
The high-latitude SNR candidate G116.6-26.1, originally found in the \textit{SRG}/eROSITA survey has been identified in the LOFAR-DR2 data with the shell-like morphology that traces the X-ray boundary of the SNR. 
The peak surface brightness extrapolated to 1~GHz is extremely low $\sim 10^{-23} \, {\rm W \, m^{-2}Hz^{-1} sr^{-1}}$, placing G116.6-26.1 well below other objects in the $\Sigma-D$ diagram. 
We argue that such  low surface brightness can be explained by the compression of relativistic electrons and magnetic fields (van der Laan model) in the Milky Way halo. In this model, the required volume emissivity in the LOFAR band of the halo gas upstream of the shock is  $\sim 10^{-42}\,{\rm Wm^{-3}Hz^{-1} sr^{-1}}$. This value is comparable to those envisaged by models postulating the existence of very extended (more than 10~kpc) synchrotron-emitting diffuse halo around the Milky Way, demonstrating that (admittedly rare) SNR in the halo can be used to probe not only the thermal gas but also the non-thermal components of the ISM.



\section*{Acknowledgments}
We are grateful to Timothy Shimwell for the discussion of the LOFAR data. 

This work is partly based on observations with the eROSITA telescope onboard \textit{SRG} space observatory. The \textit{SRG} observatory was built by Roskosmos in the interests of the Russian Academy of Sciences represented by its Space Research Institute (IKI) in the framework of the Russian Federal Space Program, with the participation of the Deutsches Zentrum für Luft- und Raumfahrt (DLR). The eROSITA X-ray telescope was built by a consortium of German Institutes led by MPE, and supported by DLR. The \textit{SRG} spacecraft was designed, built, launched, and is operated by the Lavochkin Association and its subcontractors. The science data are downlinked via the Deep Space Network Antennae in Bear Lakes, Ussurijsk, and Baikonur, funded by Roskosmos. The eROSITA data used in this work were converted to calibrated event lists using the eSASS software system developed by the German eROSITA Consortium and analysed using proprietary data reduction software developed by the Russian eROSITA Consortium.

LOFAR is the Low Frequency Array designed and constructed by ASTRON. It has observing, data processing, and data storage facilities in several countries, which are owned by various parties (each with their own funding sources), and which are collectively operated by the ILT foundation under a joint scientific policy. The ILT resources have benefited from the following recent major funding sources: CNRS-INSU, Observatoire de Paris and Université d'Orléans, France; BMBF, MIWF-NRW, MPG, Germany; Science Foundation Ireland (SFI), Department of Business, Enterprise and Innovation (DBEI), Ireland; NWO, The Netherlands; The Science and Technology Facilities Council, UK; Ministry of Science and Higher Education, Poland; The Istituto Nazionale di Astrofisica (INAF), Italy.

IK acknowledges support by the COMPLEX project from the European Research Council (ERC) under
the European Union’s Horizon 2020 research and innovation program grant agreement ERC-2019-AdG 882679.



\section*{Data availability}
X-ray data analysed in this article were used by permission of the Russian \textit{SRG}/eROSITA consortium. The data will become publicly available as a part of the corresponding \textit{SRG}/eROSITA data release along with the appropriate calibration information. The LOFAR-DR2 data are publicly available.


\bibliographystyle{mnras}
\bibliography{references} 







\bsp	
\label{lastpage}
\end{document}